\documentclass{acmsiggraph}

\title{3D Printed Stencils for Texturing Flat Surfaces}

\author{Vaibhav S. Vavilala\\Computer Graphics Lab, Columbia University}
\pdfauthor{Vaibhav S. Vavilala}


\setcopyright{none}
\begin{document}

%%% This is the ``teaser'' command, which puts an figure, centered, below 
%%% the title and author information, and above the body of the content.

 \teaser{
   \includegraphics[width=6.8in]{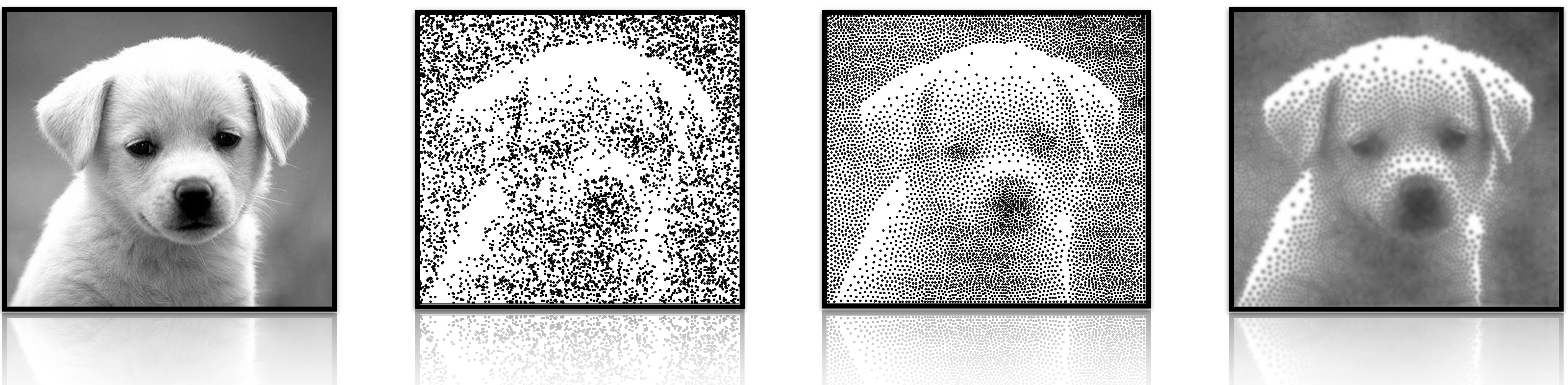}
   \caption{Our pipeline starting from an opacity channel to 3D printed stencil. Simulated spray-painted result shown on the right.}
 }

\maketitle

\begin{abstract}

We address the problem of texturing flat surfaces by spray-painting through 3D printed stencils. We propose a system that (1) decomposes an image into alpha-blended layers; (2) computes a stippling given a transparency channel; (3) generates a 3D printed stencil given a stippling and (4) simulates the effects of spray-painting through the stencil. 

\end{abstract}

%\begin{CCSXML}
%<ccs2012>
%<concept>
%<concept_id>10010147.10010371.10010382</concept_id>
%<concept_desc>Computing methodologies~Image manipulation</concept_desc>
%<concept_significance>500</concept_significance>
%</concept>
%<concept>
%<concept_id>10010147.10010371.10010382.10010236</concept_id>
%<concept_desc>Computing methodologies~Computational photography</concept_desc>
%<concept_significance>300</concept_significance>
%</concept>
%</ccs2012>
%\end{CCSXML}

%\ccsdesc[500]{Computing methodologies~Image manipulation}
%\ccsdesc[300]{Computing methodologies~Computational photography}


% The next three commands are required, and insert the user-generated keywords, 
% The CCS concepts list, and the rights management text.
% Please make sure there is a blank line between each of these three commands.

\keywordlist

%\conceptlist

%\printcopyright

\section{Introduction}

\begin{figure}[ht]
  \centering
  \includegraphics[width=3.2in]{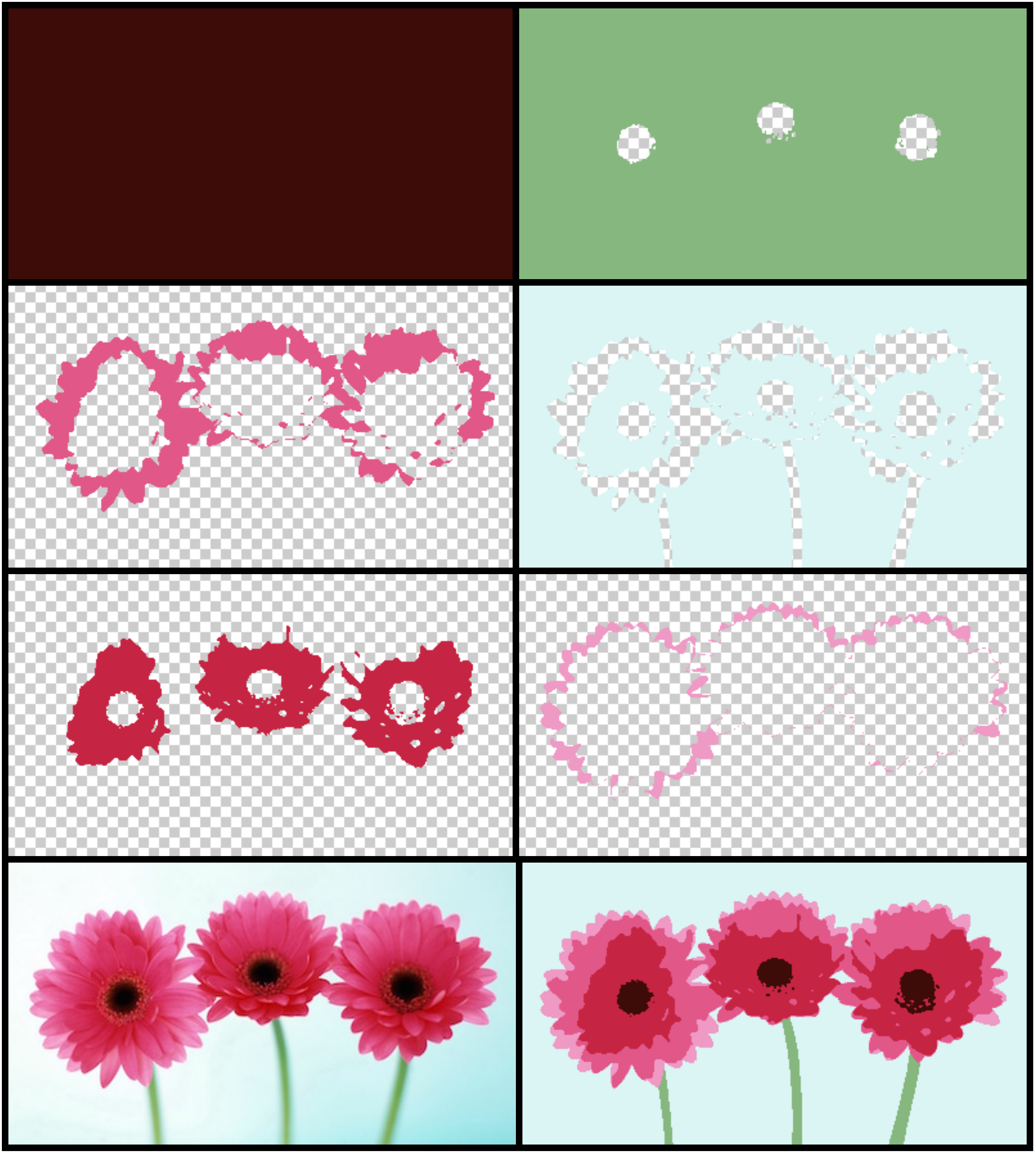}
  \caption{Hard-stencil generation for a flower (bottom left). Composite shown on bottom right.}
  \label{fig:hardStencil}
\end{figure}

Stencils have long been used for decorating walls, fabrics, and even food. Application of stencils involves placing a thin sheet against a flat surface, and spray-painting through the holes in the sheet. These stencils have been generated by hand as well as using image editing applications like Adobe Photoshop. Recently, the generation of stencils has been automated ~\cite{stencil1}. By segmenting images into piecewise constant layers using the Graph Cut Algorithm and fixing the topology~\cite{topoCut}, an arbitrary image can be decomposed into a series of stencils, with one color assigned per layer. We implement this system in MATLAB (see Figure~\ref{fig:hardStencil}). We observe that such stencils are limited to hard-boundaries and only a finite number of colors; the final composite cannot capture the millions of colors present in the input image. 

%\begin{table}[ht]
%  \centering
%  \caption{A simple table.}
%  \begin{tabular}{|r|l|}
%    \hline
%    7C0 & hexadecimal \\
%    3700 & octal \\ \cline{2-2}
%    11111000000 & binary \\
%    \hline \hline
%    1984 & decimal \\
%    \hline
%  \end{tabular}
%\end{table}
  
Here, we propose using 3-D printed stencils to produce richer composites more faithful to the original image. Our stencils can reproduce the spectrum of $\alpha$ values, as opposed to only $\alpha = 0$ or $1$. Elevation of the stencil away from the surface enables a blend between holes drilled into the stencil. 

\section{Procedure \& Results}

\begin{figure*}
  \centering
  \includegraphics[width=6.7in]{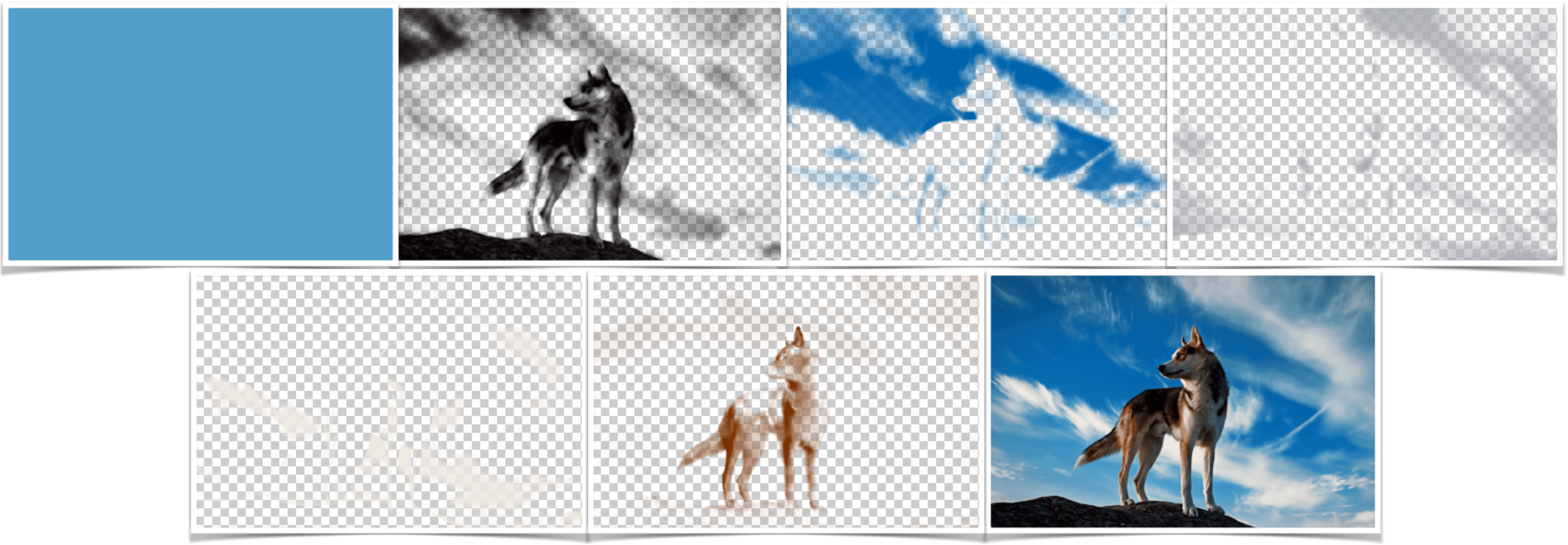}
  \caption{Computing six soft-segmented layers. Blending these layers produces a chromatically rich composite.}
  \label{fig:soft}
\end{figure*}
\begin{figure*}[h]
  \centering
  \includegraphics[width=6.7in]{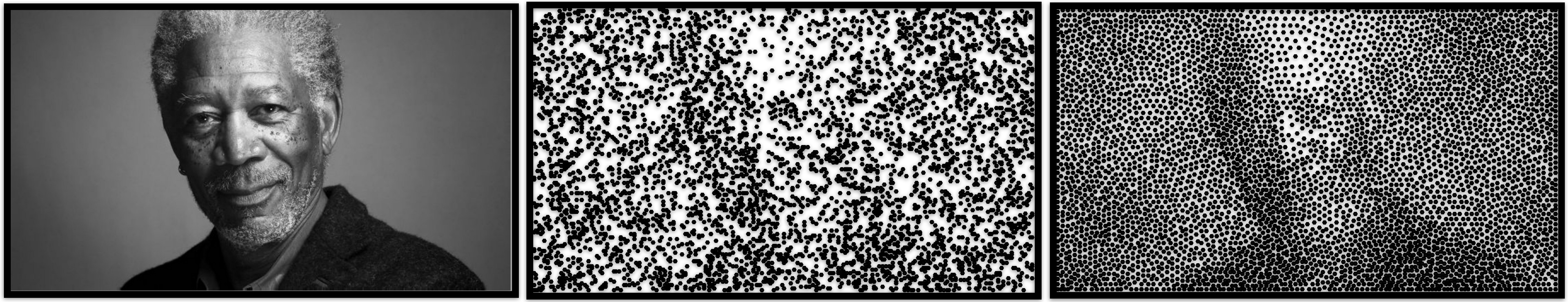}
  \caption{(Left) Input image. (Center) Importance sampling result. (Right) Result after Lloyd's relaxation. $N=13.9K$ dots.}
  \label{fig:stip}
\end{figure*}

Our method takes as input an image and outputs a sequence of 3D printed stencils. We first decompose the input image into a user-specified number of layers. We then produce stipplings of each layer. We finally engineer a 3D model corresponding to each stippling layer. Our contribution also includes a spray-paint simulator that will predict the effects of spray-painting through an arbitrary stencil. 

%\begin{equation}
%\label{eq:j}
%J_{ij} = \frac{1}{N}\sum_{\mu=1}^p \xi^\mu_i\xi^\mu_j= J_{ji}\hspace{1cm} i\neq j.
%\end{equation}

\subsection{Image Decomposition}
We outline our method for decomposing images into layers, that when alpha blended, produce a composite as close to the original as possible. We rely on the Porter-Duff over operator for our compositions~\cite{over}, defined as: 

\begin{equation}
\label{eq:Porter}
C_i = (1-\alpha_{i}^{k})C_{i-1} + c_i\alpha_{i}^{k}
\end{equation}

where there are $i=1...L$ layers, $C_i$ is the composite pixel color after $i$ layers, $c_i$ is the color associated with layer $i$, and $\alpha_{i}^{k}$ is the opacity at pixel $k$ and layer $i$. We also compute a background color for $i=0$ where $\alpha=1$ at every pixel. This background layer can be interpreted as "priming" the surface. 

The image decomposition problem aims to compute $L$ layers of $\alpha$ values and $L+1$ colors associated with these layers. We define an energy function that takes into account the attributes we'd like to be true of a solution.

\textit{Data Term} We minimize the difference between composite and input pixel value as follows: 
	\begin{equation}
	\label{eq:Edata}
	E_{data} = \sum_{p = 1}^P \big(X_{p}^{R} - Y_{p}^{R}\big)^2 + \big(X_{p}^{G} - Y_{p}^{G}\big)^2 + \big(X_{p}^{B} - Y_{p}^{B}\big)^2 
	\end{equation}
	where $X$ is the input image, $Y$ is the composite, and there are $P$ pixels. The three terms correspond with three color channels, $(R,G,B)$.

\textit{Smoothness Term} Our smoothness term aims for smoothly varying transparency values in each layer. Since we are blending adjacent stippling dots, local sharp changes in transparency are difficult to reproduce. Let $Y(1), Y(2),...,Y(L)$ be the computed $\alpha$ channels. We define our smoothness term as follows: 
	\begin{equation}
	\label{eq:Esmooth}
	E_{smooth} = \sum_{l = 1}^L \sum_{p = 1}^P \sum_{n \in N_8(p)} \big(Y_p(l) - Y_n(l)\big)^2 
	\end{equation}
	where $N_8(p)$ refers to the eight neighbors of pixel $p$. 

\textit{Sparsity Term}
	We begin with the presupposition that it is easier to replicate painted regions with $\alpha=1$ or $\alpha=0$, and there will be more error associated with intermediate values. We introduce a sparsity term to drive transparency values to these extremes by penalizing intermediate values. Driving the sum of the $\alpha$ values among the top $L$ layers to 1, independently for each pixel, achieves this goal. 
	\begin{equation}
	\label{eq:Esparse}
	E_{sparse} = \sum_{p = 1}^P \bigg(1-\sum_{l = 1}^L Y_{p}(l)\bigg)^2 
	\end{equation}

We then set up a convex optimization problem, solving for $L$ layers sequentially, then solving for all the colors at once, and repeating until convergence. If we let $\gamma_{data}, \gamma_{smooth}$, and $ \gamma_{sparse}$ tune the relative strength of the energy terms, the final objective function is:

	\begin{equation}
	\label{eq:objective}
	min. \big(\gamma_{data}E_{data} + \gamma_{smooth}E_{smooth} + \gamma_{sparse}E_{sparse}\big)
	\end{equation}

Given a target number of layers, K-means and quantization are valid methods for generating initial color guesses for each layer. Then, we can assign each pixel to the nearest color among the $L$ choices to initialize the layers. These layers and colors are then modified by our sequential quadratic programming with blocked coordinate descent method.

\subsection{Stippling}

\begin{figure*}[h]
  \centering
  \includegraphics[width=6.7in]{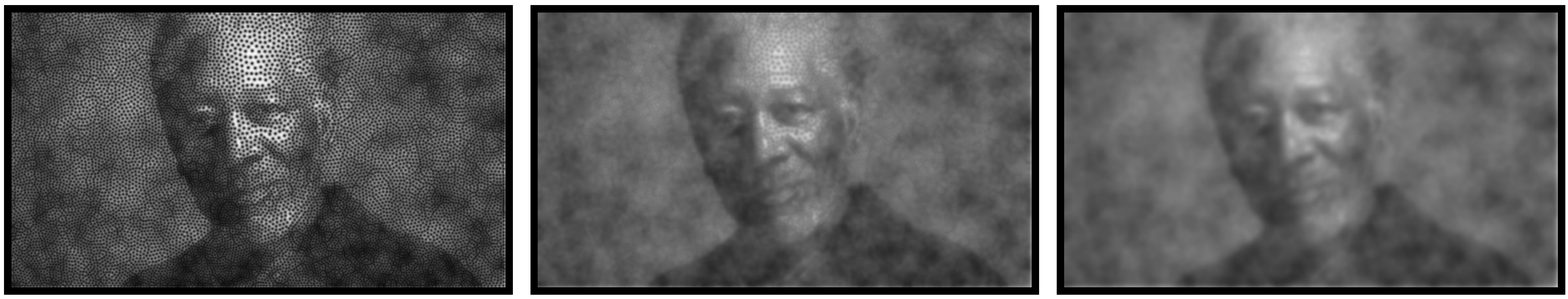}
  \caption{Spray-paint simulation with (Left) $h=2$, (Center) $h=7$ (Right) $h=15$.}
  \label{fig:morgan_spray}
\end{figure*}

\begin{figure*}[h]
  \centering
  \includegraphics[width=6.7in]{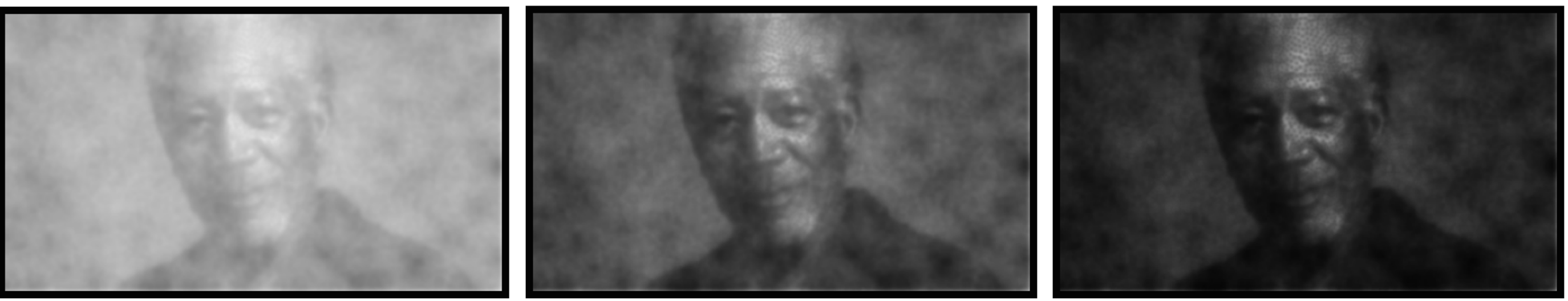}
  \caption{Effects of spray-painting for increasing amounts of time.}
  \label{fig:morganInk}
\end{figure*}

\begin{figure*}[h]
  \centering
  \includegraphics[width=6.8in]{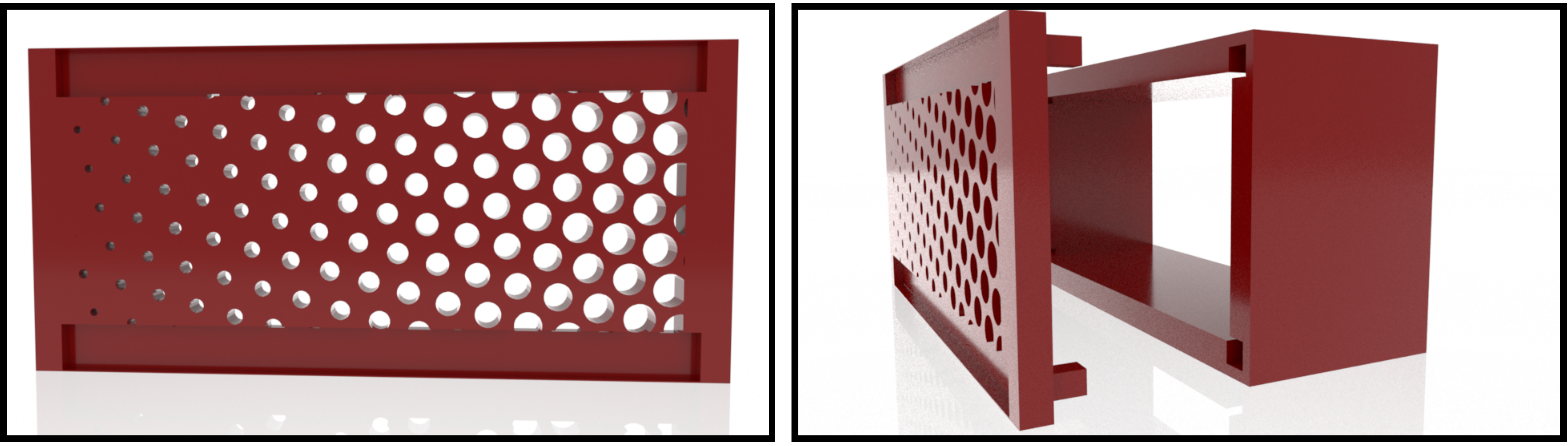}
  \caption{Renders of our 3D printed stencils showing a sample stippling pattern. The stencil and wall fit like lock and key, so walls of different heights can be applied to the same stencil.}
  \label{fig:3Dstencil}
\end{figure*}

%\begin{figure*}[h]
  %\centering
  %\includegraphics[width=6.8in]{images/futureWork}
  %\caption{Simulated composites of our alpha-blended layers. Errors within each layer augment in the final composition. Future work involves enhancing the %pipeline to improve composition quality. }
  %\label{fig:FutureWork}
%\end{figure*}

Stippling is an artistic technique whereby varying degrees of shading are approximated by small dots. A common approach to generating quality stippling involves employing Lloyd's relaxation on a weighted centroidal Voronoi tessellation~\cite{stipple}. We implement such a method here. To initialize the seed points, we use importance sampling weighted by opacity. Importance sampling, unlike dithering, avoids aliasing artifacts that Lloyd's relaxation cannot always fix. 

%\begin{figure}[ht]
%  \centering
%  \includegraphics[width=3.0in]{images/ferrari_laferrari}
%  \caption{Ferrari LaFerrari. Image courtesy Flickr user ``gfreeman23.''}
%  \label{fig:ferrari}
%\end{figure}

\subsection{Spray-Paint Simulator}
Previous work shows that the distribution of ink expelled from a spray-paint nozzle is approximately Gaussian~\cite{romain}. We treat each point in the CVT as a 2-D Gaussian with its center as the mean, and variance given by $\sigma^2 = \frac{1}{4}r^2(1+h)$. We fix the radius to be $0.05 cm$, a reasonable lower bound on the smallest hole that can be drilled through a 3-D printed material ~\cite{shape}. We treat $h$, the height the nozzle lies away from the material, as a tunable parameter. For simplicity, $h$ is the same for all dots in our simulations. Tuning the Gaussian prefactor simulates the effects of time; the greater the coefficient, the more time spent painting through each hole. 

Let $\mathcal{N}_1,\mathcal{N}_2,...,\mathcal{N}_N$ be the set of Gaussians, where there are $N$ dots. A successful simulation must take into account that the intensity of ink at any pixel is capped at $1$; adding more dots near a pixel or spray-painting for more time cannot push the maximum opacity value above $\alpha=1$. These bounds on the pixel intensity motivate using complementary probability for our simulation. We produce the simulated result as follows:

	\begin{equation}
	\label{eq:sim}
	I(x,y) = 1 - \prod_{i=1}^{N}\Big(1-\min\big[\mathcal{N}_i(x,y),1\big]\Big)
	\end{equation}

where $\mathcal{N}_i(x,y)$ means evaluating the height of Gaussian $i$ at pixel site $(x,y)$. We apply equation~\ref{eq:sim} on all pixels to compute the predicted spray-painted result. 

\subsection{3D Model}
Starting from our stippled layer, we extrude a stencil and surrounding wall with tunable height~\cite{gp}. Functionally, the wall elevates the stencil above the surface; it also prevents unintended paint from leaving the target region. The output of this step is a 3D-printable mesh.  
%\section{Results}
%
%Here we show results illustrating our method. 

%N=13.9K dots
\section{Conclusion}
We have presented a pipeline to bring 3D printing to surface decoration. Future work would include printing and spray-painting the physical stencils, and comparing the composites with the input and simulated images. We also expect that additional work on the stippling algorithm would result in a more faithful simulated result. 

\section*{Acknowledgements}

The author is grateful to Alec Jacobson and Changxi Zheng for their generous mentorship, as well as Eitan Grinspun for helpful conversations. 

This work was completed 05/15 - 12/15 and programmed in MATLAB. 

\bibliographystyle{acmsiggraph}
\nocite{*}
\bibliography{template}
\end{document}